%% file: main.tex
\documentclass[conference]{IEEEtran}
\IEEEoverridecommandlockouts
\usepackage{cite}
\usepackage{amsmath,amssymb,amsfonts}
\usepackage{graphicx}
\usepackage{textcomp}
\usepackage{xcolor}

\usepackage{cuted}
\usepackage{caption}
\usepackage{float}
\usepackage{lipsum}
\usepackage{changepage}
\usepackage{hyperref}

\usepackage{booktabs}
\usepackage{tabularx}

\usepackage[ruled,vlined,linesnumbered]{algorithm2e}

\usepackage{listings}
\def\BibTeX{{\rm B\kern-.05em{\sc i\kern-.025em b}\kern-.08em
    T\kern-.1667em\lower.7ex\hbox{E}\kern-.125emX}}
\begin{document}

\title{VeriGraphi: A Multi-Agent Framework of Hierarchical RTL Generation for Large Hardware Designs\\
}

\author{\IEEEauthorblockN{1\textsuperscript{st} Sazzadul Islam}
\IEEEauthorblockA{\textit{Dept. of Computer Science \& Eng.} \\
\textit{University of South Florida}\\
Tampa, Florida, United States \\
sazzadulislam@usf.edu}

\and

\IEEEauthorblockN{2\textsuperscript{nd} Tasnim Tabassum}
\IEEEauthorblockA{\textit{Dept. of Computer Science \& Eng.} \\
\textit{University of South Florida}\\
Tampa, Florida, United States \\
tasnimtabassum@usf.edu}

\and

\IEEEauthorblockN{3\textsuperscript{rd} Hao Zheng}
\IEEEauthorblockA{\textit{Dept. of Computer Science \& Eng.} \\
\textit{University of South Florida}\\
Tampa, Florida, United States \\
haozheng@usf.edu}
}

\maketitle

\begin{abstract}
Generating synthesizable Verilog for large, hierarchical hardware designs remains a significant challenge for large language models (LLMs), which struggle to replicate the structured reasoning that human experts employ when translating complex specifications into RTL. When tasked with producing hierarchical Verilog, LLMs frequently lose context across modules, hallucinate interfaces, fabricate inter-module wiring, and fail to maintain structural coherence—failures that intensify as design complexity grows and specifications involve informal prose, figures, and tables that resist direct operationalization. To address these challenges, we present VeriGraphi, a framework that introduces a spec-anchored Knowledge Graph as the central architectural substrate driving the RTL generation pipeline. VeriGraphi constructs a Hierarchical Design Architecture (HDA), a structured knowledge graph that explicitly encodes module hierarchy, port-level interfaces, wiring semantics, and inter-module dependencies as first-class graph entities and relations. Built through iterative multi-agent analysis of the original specification, this Knowledge Graph provides a deterministic, machine-checkable structural scaffold before code generation. Guided by the KG, a progressive coding module incrementally generates pseudo-code and synthesizable RTL while enforcing interface consistency and dependency correctness at each submodule stage. We evaluate VeriGraphi on a benchmark comprising three representative specification documents published by the National Institute of Standards and Technology (NIST) and their corresponding implementations, and we present a RISC-V 32I processor as a detailed case study to illustrate the full pipeline. The results demonstrate that VeriGraphi enables reliable hierarchical RTL generation with minimal human intervention for RISC-V, marking a significant milestone for LLM-generated hardware design while maintaining strong functional correctness.
\end{abstract}


\input{introduction}

\input{related_work}
\input{framework}
\input{experiments}
\input{results}

\input{conclusion_futureWork}
\input{acknowledgement}

\bibliographystyle{IEEEtran}
\bibliography{references}


\end{document}

%% file: introduction.tex
\section{Introduction}

The rapid growth of large language models (LLMs) has significantly accelerated the adoption of artificial intelligence (AI) across diverse application domains, leading to an increasing imbalance between the pace of AI software innovation and the rate of underlying hardware advancement. While hardware remains the fundamental enabler of scalable and energy-efficient AI systems, progress in hardware design automation continues to lag due to the intrinsic complexity of hardware architectures and the heavy reliance on human expertise. As a result, scalable chip design has emerged as a critical bottleneck in supporting the momentum of AI-driven innovation.

Recent advances show that state-of-the-art LLMs can automate substantial portions of the software development pipeline, including the generation of syntactically correct and semantically meaningful code in high-level programming languages such as Python and C++, even for modular and object-oriented designs\cite{hou2024largelanguagemodelssoftware, rahman2025syntheticbenchmarksevaluatingllm}. Conversely, applying LLMs to hardware description language (HDL) generation --particularly Verilog---remains substantially more challenging. Hardware designs are inherently hierarchical, structurally constrained, and timing-sensitive, requiring precise coordination among modules, interfaces, and signals. LLMs struggle to reliably generate complex hierarchical Verilog while maintaining both syntactic correctness and functional fidelity, especially for rarely implemented or highly specialized designs\cite{tang2025hivegen,yang2025largelanguagemodelverilog,yu2025spec2rtl}. Consequently, human intervention is often required to resolve hallucinations, misaligned hierarchies, and incorrect wiring decisions.

A fundamental limitation arises from the gap between the unstructured nature of real-world hardware specification documents and the requirements of HDL generation. Specifications are typically lengthy PDF documents that contain informal text, tables, figures, and equations, making it difficult for LLMs to directly extract and operationalize design intent\cite{yu2025spec2rtl}. Flat prompting further exacerbates this issue, frequently causing LLMs to lose context, fabricate missing structural details, or incorrectly infer module connectivity when tasked with generating Verilog directly from specifications.

To address these challenges, recent research has explored hierarchical prompting and agent-based multi-stage generation pipelines for LLM-assisted hardware design, aiming to decompose complex specifications into manageable submodules and iteratively refine RTL implementations \cite{nakkab2024rome, tang2025hivegen, yu2025spec2rtl}. Beyond hierarchical decomposition, emerging efforts investigate graph-based representations to explicitly model architectural entities and relationships within specifications or RTL designs, enabling systematic reasoning for tasks such as assertion generation, architecture-to-RTL translation, and neighborhood-aware retrieval \cite{bai2025assertionforge, yang2025automating, verirag-aspdac}. In parallel, retrieval-augmented and summarization-oriented systems, such as NotebookLM, demonstrate strong capability in extracting and organizing information from Verilog specifications; however, these tools primarily support document understanding rather than synthesis-oriented, agentic RTL generation \cite{google_notebooklm_2024}. Although these directions collectively advance LLM-assisted hardware design, challenges remain in achieving tightly integrated structural representations that ensure deterministic hierarchy construction, interface consistency, and end-to-end traceability within the generation pipeline.

Motivated by these limitations, we propose a structured knowledge-graph–driven pipeline that serves as a unifying architectural substrate across specification analysis and RTL generation. Instead of relying on text-centric plans or post-hoc structural recovery, our approach constructs a spec-anchored knowledge graph that explicitly captures hierarchy, interfaces, and inter-module relationships prior to code generation. This explicit hierarchy breaks a complex RTL task into smaller module-level problems that are easier to generate and revise for LLMs, while also supporting bottom-up verification for earlier bug identification and clearer fault localization before full-system integration. At the same time, the resulting structured representation enables deterministic integration and consistent interface modeling by reducing implicit structural inference during generation. Furthermore, the knowledge graph supports organized reasoning and cross-stage traceability as the design evolves. In this manner, the knowledge graph functions as an LLM-friendly architectural plan within an end-to-end pipeline.

The main contributions of this work are summarized below:

\begin{itemize}
    \item We propose a spec-anchored \textbf{Knowledge Graph (KG)}–driven RTL generation framework integrating architectural analysis, hierarchical modeling, and progressive coding within a unified agentic pipeline, where the KG is an intermediate representation encoding modules, ports, signals, and hierarchical relations through explicit semantic edges, enabling module-level validation and deterministic structural integration prior to RTL generation.
    
    \item We design a KG-guided progressive coding mechanism that incrementally generates pseudo-code and synthesizable RTL while enforcing dependency correctness, and structural alignment with the graph model.
    
    \item We evaluate \textbf{VeriGraphi} through both synthesizability and functional correctness, using specification-driven testbenches for module-level and integration-level validation. Our results show successful generation of complex designs, including RV32I with minimal human intervention and HMAC with zero intervention, while achieving significantly fewer coding iterations, and thus lower generation cost, than prior work.
\end{itemize}

%% file: related_work.tex
\section{Related work}

\subsection{LLM-Driven RTL Generation Pipelines}

Prior work has explored hierarchical and agent-based prompting strategies for LLM-assisted hardware design. ROME \cite{nakkab2024rome} introduced hierarchical prompting for decomposing large designs into submodules and demonstrated one of the earliest LLM-generated processors. However, its human-driven hierarchical prompting (HDHP) mode depends on expert-provided decompositions, while its purely generative hierarchical prompting (PGHP) mode is fragile for rarely implemented designs; in both cases, top-level wiring and interface integration remain largely dependent on the LLM. HiVeGen \cite{tang2025hivegen} improves scalability through hierarchical generation, module reuse, and runtime structural correction, but it still relies on predefined templates and application-oriented inputs such as LLVM-based extraction from C/C++ kernels, which limits direct use on arbitrary PDF specifications. ChatCPU \cite{chatcpu} demonstrates strong results for CPU design by combining fine-tuned LLMs with a processor description language and verification flow, but it is tightly coupled to predefined CPU templates and a domain-specific specification format. Spec2RTL-Agent \cite{yu2025spec2rtl} further advances automation through a multi-agent workflow for spec analysis, code generation, and debugging, yet its intermediate plan remains text-centric rather than machine-checkable, making structural verification and cross-stage error attribution difficult. Overall, existing RTL-generation pipelines improve decomposition and refinement, but still face common limitations in deterministic hierarchy construction, explicit interface modeling, and reliable top-level integration.

\subsection{Graph-Based \& Retrieval-Augmented Hardware Reasoning}

AssertionForge \cite{bai2025assertionforge} introduces a knowledge-graph workflow for formal verification by extracting entities from natural-language specifications and refining them with structural information from existing RTL. This improves assertion generation, but its KG construction assumes RTL is already available and is aimed at verification-time context synthesis rather than spec-first RTL generation. VeriRAG \cite{verirag-aspdac} combines hardware knowledge graphs with vector retrieval to provide symbolic and semantic context for Verilog and SVA generation, demonstrating the value of retrieving structured neighborhoods instead of injecting flat prompt contexts. However, its graph is primarily built from an RTL corpus and depends on curated retrieval coverage, which can weaken traceability to original specification intent and become brittle when retrieval misses rare or novel modules. Yang et al. \cite{yang2025automating} use a neural-symbolic blueprint graph to translate unstructured architectural descriptions into symbolic specifications for RTL generation and verification, showing the benefit of structured intermediate representations over flat text. However, their blueprint graph mainly captures high-level architectural structure and behavior, with limited modeling of detailed port-level connectivity and wiring constraints needed for deterministic RTL integration.

%% file: framework.tex
\section{Framework}
\begin{figure*}[t]
    \centering 
    \includegraphics[width=\textwidth]{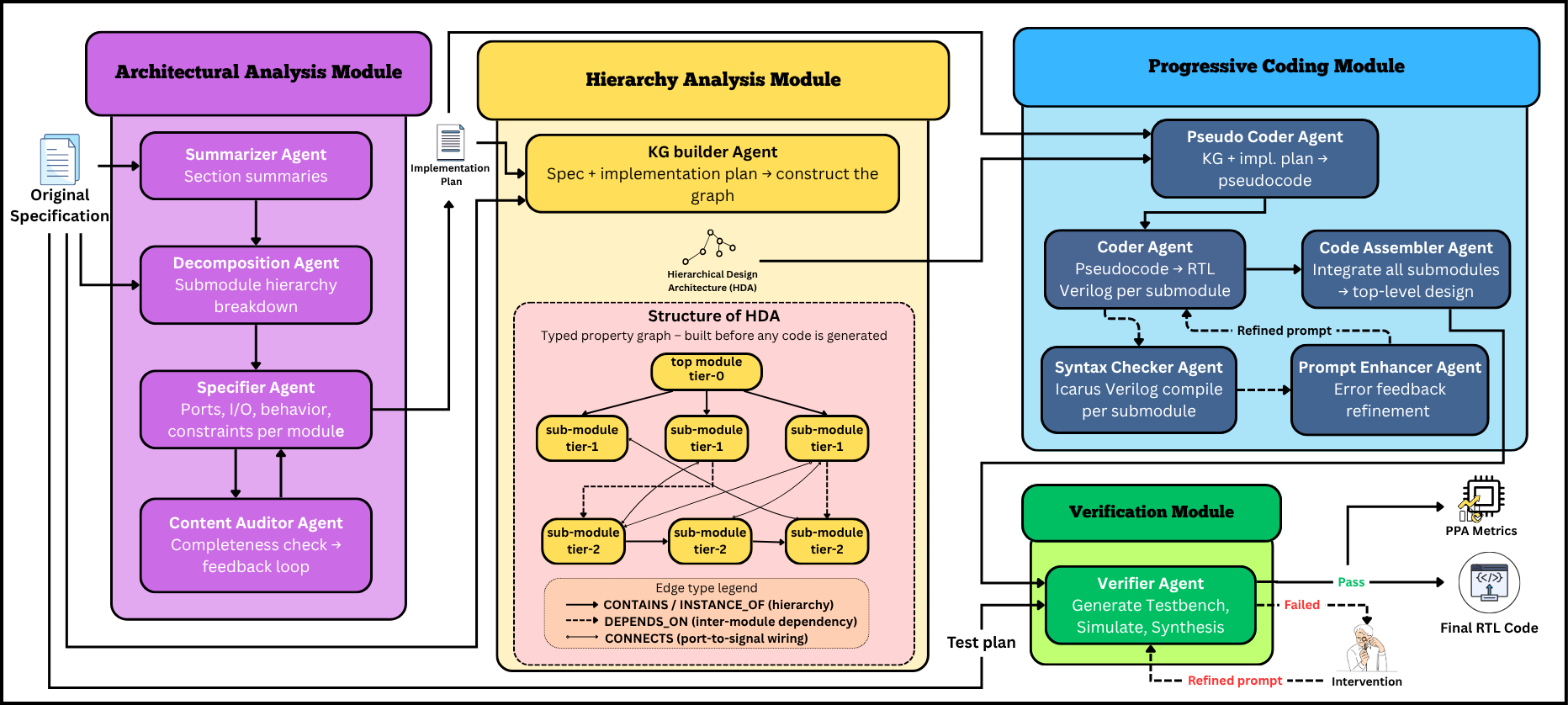}
    \caption{System Architecture.}
    \label{fig_framework}
\end{figure*}

\subsection{Overview}
To achieve an automated hardware design pipeline directly from raw specs, we propose a structured architecture, illustrated in Fig. \ref{fig_framework}, composed of three tightly coupled modules. Each module incrementally transforms specification-level intent into a synthesizable Verilog design, while maintaining an organized Knowledge Graph (KG) that serves as an explicit architectural representation. The KG is constructed based on the decomposed architecture and functional implementation plan to enable systematic reasoning, traceability, and deterministic module integration. The framework includes:

\subsubsection{\textbf{Architectural Analysis Module}}
This module performs an organized interpretation of the specification document, emulating the reasoning process of a hardware expert. It analyzes and summarizes the document, decomposes the design into functional submodules, and formalizes their roles, interfaces, and dependencies. The output is a detailed implementation plan that defines the architectural blueprint for later stages.

\subsubsection{\textbf{Hierarchy Analysis Module}}
Building upon the implementation plan, this module constructs and updates the Knowledge Graph. It captures the Device Under Test (DUT) interface (including ports, clock/reset conventions, parameters, and constraints) and models explicit hierarchical relationships among submodules. Wiring semantics, interface mappings, and structural dependencies are represented as typed graph entities and relations, ensuring deterministic hierarchy formation and machine-checkable connectivity.

\subsubsection{\textbf{Progressive Coding Module}}
Guided by the implementation plan and the Knowledge Graph, this module incrementally generates synthesizable Verilog code in a progressive manner. Submodules are implemented, verified, and assembled into the top-level design through an iterative closed-loop process involving code generation, syntax verification, and prompt optimization.

\subsubsection{\textbf{Verification Module}}
This module verifies the generated RTL in a staged manner, beginning with submodule-level checking and ending with full top-module validation. It derives specification-grounded testbenches, runs simulation for both module-level and integration-level correctness, and then performs synthesis and PPA evaluation; only designs that pass both functional verification and synthesizability checks are accepted as final outputs, while failures trigger human intervention and further refinement.

Each module	is detailed in the following subsections.

\subsection{Architectural Analysis Module}
The Architectural Analysis Module is inspired by the Iterative Understanding and Reasoning paradigm introduced in Spec2RTL \cite{yu2025spec2rtl}, where the objective is to transform lengthy, unstructured, and LLM-unfriendly specification documents into a structured implementation plan. Real-world hardware specifications often contain extensive textual definition interleaved with figures, tables, and informal explanations, making direct RTL generation unreliable. This module addresses that challenge by iteratively organizing, extracting, and validating essential architectural elements to produce a well-structured, LLM-friendly implementation plan.

The module is implemented as a tightly coupled multi-agent pipeline, where specialized LLM agents collaborate to progressively refine architectural intent. The agents are described below.

\subsubsection{\textbf{Summarizer Agent}}
Processing large PDF specifications is challenging because figures and tables are often lost in raw text extraction. To address this, we use a custom preprocessing pipeline that extracts both text and visual artifacts from the specification. Text is extracted with PyPDF, while a Python-based preprocessing tool captures figures and tables as indexed image files \textit{(e.g., figure\_1, figure\_2)}, preserving their original numbering and correspondence in the document. The specification PDF and these extracted visual artifacts together form the primary system input. Since hardware specifications are typically organized into sections, the \textit{Summarizer Agent} generates structured summaries at section granularity. During summarization, it reasons over the text to detect implicit or explicit figure and table references, retrieves the corresponding pre-extracted images, and incorporates them into the reasoning process. This multimodal grounding helps capture structural and architectural details conveyed through diagrams. The output is a structured JSON summary for each section, including metadata such as referenced figures and tables, which serves as an LLM-friendly multimodal context for downstream decomposition and architectural analysis.

\subsubsection{\textbf{Decomposition Agent}}
Complex hardware systems inherently follow hierarchical design principles. The Decomposition Agent takes both the section-wise summaries and the original specification content as input to systematically identify all required sub-functions or submodules necessary to implement the target top-level design. Beyond identifying components, the agent determines a correct implementation order by explicitly modeling inter-module dependencies, producing an ordered and dependency-aware decomposition. When the specification does not explicitly define a hierarchy, the LLM is guided to infer a plausible hierarchical structure based on general architectural knowledge, which is then exposed for optional human review and refinement. A key aspect of this agent is its ability to associate each submodule with precise \textit{spec\_reference} links to relevant sections of the original document. This allows subsequent agents to operate on focused, context-specific excerpts rather than the full specification, significantly reducing token usage and improving efficiency. The output is a hierarchical structural blueprint consisting of submodules, their functional roles, dependencies, and targeted specification references, aligned with the original design intent.

\subsubsection{\textbf{Specifier Agent}}
Given the identified submodules and their associated references, the \textit{Specifier Agent} extracts detailed implementation-relevant information for each component. Rather than reprocessing the entire specification, it operates using the structured summaries while cross-checking with the source document to maintain fidelity. For each submodule, it formalizes inputs, outputs, functional behavior, constraints, and relevant specification anchors. The agent operates in an iterative refinement loop, updating its descriptions based on feedback from the \textit{Content Auditor Agent} to ensure completeness and consistency. The output is a structured description suitable for downstream graph construction.

\subsubsection{\textbf{Content Auditor Agent}}
The \textit{Content Auditor Agent} reviews the generated descriptions to ensure completeness, consistency, and alignment with the specification. It evaluates each submodule description across four key dimensions: structural completeness \textit{(e.g., presence of all required fields such as inputs/outputs with data types and bit widths)}, correctness with respect to the specification, clarity of implementation steps, and readiness for RTL coding. For instance, if a description omits signal widths, provides ambiguous step ordering, or references non-specific sections, the agent flags these as issues and suggests concrete corrections. Based on this evaluation, the agent produces structured feedback indicating whether the description is valid, along with identified issues and actionable suggestions. Descriptions that fail validation are routed back to the \textit{Specifier Agent} for refinement, while valid ones may still receive improvement suggestions. This feedback-driven loop ensures that each submodule description is sufficiently precise, unambiguous, and implementation-ready before proceeding to downstream structural modeling.

After completing these stages, the system produces a comprehensive and hierarchically organized implementation plan for the top module and its constituent submodules. This structured plan serves as the foundation for the subsequent Hierarchy Analysis Module, where explicit Knowledge Graph representations are constructed from both the original specification and the refined implementation plan.

\subsection{Hierarchy Analysis Module}

A major limitation in prior LLM-based hardware generation is the lack of an explicit and specifiable representation of connections between submodules when composing a top-level design. In our experiments, we observed that accurate, detailed connectivity becomes crucial as designs grow in complexity and depend on multi-layer hierarchical child modules. When such detail is missing, LLMs frequently hallucinate I/O ports, internal signals, and wiring, leading to integration failures even when individual submodules appear correct.

To address this, our framework introduces a \textit{KG Builder Agent} within the Hierarchy Analysis Module. It identifies the required modules, captures their dependencies, organizes their hierarchical relationships, and records the structural information needed for downstream code generation and assembly.

\subsubsection{\textbf{Hierarchical Design Architecture (HDA)}}

The \textit{Hierarchical Design Architecture (HDA)} is a design-session knowledge graph constructed by the \textit{KG Builder Agent} to represent the target design as an explicit, machine-readable hierarchy with well-defined wiring semantics. HDA is generated using two inputs: (1) the structured implementation plan produced by the Architectural Analysis Module, and (2) the original spec including section and figure references. The graph is implemented as a property graph consisting of nodes and typed relationships with associated key--value properties.

\paragraph{\textbf{Construction Workflow (Spec-Grounded)}}
For each module required in the target design, the \textit{KG Builder Agent} follows a specification-grounded construction process.

\textbf{Graph construction from implementation plan.}
The agent builds the HDA using the structured implementation plan together with the original specification. Based on the decomposed modules and their dependencies, it creates module nodes and establishes hierarchical relationships.

\textbf{Explicit connectivity and constraint modeling.}
Ports, signals, parameters, and wiring constraints are explicitly defined using information grounded in the specification and implementation plan, rather than being inferred during code generation. This ensures consistent and unambiguous inter-module connections. The resulting HDA forms a complete and deterministic architectural scaffold capturing both hierarchy and connectivity, which guides downstream RTL generation and reduces hallucinated interfaces or incorrect wiring.

\paragraph{\textbf{Graph Semantics}}
HDA models the target design at module granularity while making both hierarchy and connectivity explicit.

\textbf{Nodes.}
Each node represents a module in the architecture. Module nodes are annotated with functional descriptions, interface definitions (ports and widths), configuration parameters, and optional references to specification sections or figures.

\textbf{Edges.}
Edges show structural relationships among modules:
\begin{itemize}
\item \textbf{CONTAINS / INSTANCE\_OF}: hierarchical relationships describing which module calls which submodule.
\item \textbf{DEPENDS\_ON}: defining the inter-module dependencies.
\item \textbf{CONNECTS}: wiring relationships describing port-to-signal connections across modules.
\end{itemize}

By explicitly encoding both hierarchy and connectivity, the HDA provides a structural blueprint that constrains LLM-driven RTL generation and reduces the likelihood of hallucinated interfaces or wiring errors during module integration. The constructed HDA is also saved as a JSON file, together with reference links to the original specification sections, and visualized graphically to enable optional human inspection without tracing hierarchy and connectivity manually from the full document.

\subsection{Progressive Coding Module}\label{subsec:framework_progressive_coding_module}
With the implementation plan from the Architectural Analysis Module and the knowledge graph from the Hierarchy Analysis Module, the pipeline proceeds to the code generation stage, referred to as the \textbf{Progressive Coding Module}. This module consists of five coordinated agents that collaboratively generate compilable, hierarchical RTL code in a staged, bottom-up manner guided by the knowledge graph.
  
\subsubsection{\textbf{Pseudo Coder Agent}}
Rather than directly generating RTL, the pipeline first produces structured hardware pseudocode. This agent takes as input both the KG node (capturing ports, parameters, and interface constraints) and the corresponding implementation plan entry (capturing functionality and step-wise logic). It synthesizes these into pseudocode that explicitly defines control flow, data operations, and module behavior. This intermediate representation is critical for bridging high-level specification and RTL, allowing structured reasoning without syntax constraints, and is stored for subsequent use by the \textit{Coder Agent}.

\subsubsection{\textbf{Coder Agent}}
The Coder Agent translates the generated pseudocode into clean, synthesizable Verilog code. The agent is explicitly guided to produce synthesizer-friendly RTL compatible with standard toolchains\cite{openroad}, including compilation using \textit{iverilog (-g2012 -t null -Wall)} and downstream synthesis. This ensures that generated code adheres to practical hardware design constraints beyond syntactic correctness.

\subsubsection{\textbf{Syntax Checker Agent}}
Each generated submodule is validated using Icarus Verilog\cite{icarus_verilog}. Early-stage compilation at the submodule level ensures that errors are detected before integration, preventing propagation across hierarchical dependencies and simplifying debugging.

\subsubsection{\textbf{Prompt Enhancer Agent}}
If compilation fails, an iterative correction loop is triggered. The \textit{Prompt Enhancer Agent} refines the generation prompt using the latest full code, compiler error messages, and a compressed history of previous failures. To maintain efficiency, only erroneous lines and corresponding errors from prior iterations are retained. This significantly reduces token usage (by approximately 80\%) while preserving essential debugging context. The refined prompt is then used to regenerate the RTL via the Coder Agent. This loop continues until successful compilation or a retry limit is reached.

\subsubsection{\textbf{Code Assembler Agent}}
After all submodules are successfully generated, the \textit{Code Assembler Agent} constructs the top-level module. Using INSTANCE\_OF and CONNECTS relations from the knowledge graph, it determines module instantiations and exact port wiring to generate the hierarchical wrapper. The assembled design is then validated through syntax checking with limited retries using direct error feedback, ensuring a fully integrated and compilable RTL design.

The overall process follows a bottom-up generation strategy, where modules are ordered based on hierarchy (leaf modules first), ensuring that dependencies are resolved prior to integration. This staged approach enforces structural correctness, improves convergence efficiency, and enables reliable generation of complex hierarchical RTL designs. The pipeline for the Progressive Coding Module is shown in Algorithm~\ref{alg:coding_pipeline}.

\subsection{Verification Module}
The Verification Module is driven by a specification-based Verifier Agent.

\subsubsection{\textbf{Verifier Agent}}
To verify the RTL generated by VeriGraphi, the \textit{Verification Agent} employs Claude Sonnet 4.6\cite{anthropic2026claude} to construct a specification-driven verification pipeline in four sequential steps, grounded in the specification documents rather than the generated RTL, ensuring testbenches serve as independent ground truth verified against the spec.

\paragraph{{\textbf{Step 1 — Specification Summarization}}}
The Verification Agent is provided the full specification PDFs as project-level context. For each design, it extracts a structured per-module summary covering functional role, port-level interface, behavioral constraints, and normative requirements:

\begin{lstlisting}[basicstyle=\ttfamily\scriptsize,language=Verilog, frame=single]
    Module     : <ModuleName>
    Inputs     : signal_a [W:0], signal_b [W:0], clk
    Outputs    : result [W:0]
    Behavior   : <Functional description from spec section>
    Constraints: <Bit-width rules, valid ranges, boundary conditions>
    Spec ref   : Section X.Y
\end{lstlisting}

This summary is constructed without reference to the RTL, preventing test derivation from inheriting implementation-level bugs or assumptions.

\paragraph{{\textbf{Step 2 — Module-Level Test Case Derivation}}}
Each normative behavioral claim is mapped to a concrete, checkable condition. Boundary conditions, all-zero and all-one patterns, spec-defined constants, and known-answer pairs are extracted directly from the spec. For modules with undefined RTL functions, tests are authored against spec requirements and annotated as expected failures with documented bug references. Representative cases:

\begin{lstlisting}[basicstyle=\ttfamily\scriptsize,language=Verilog, frame=single]
// TC-MOD-01: All-zero inputs -> output must not be X
set(inputs=0); wait(N cycles); check("TC-MOD-01", output !== X);

// TC-MOD-02: Spec-defined constant -> golden-path output
set(input=<spec_constant>); ...
\end{lstlisting}

\paragraph{{\textbf{Step 3 — Integration-Level Test Construction}}}
Integration testbenches validate end-to-end data propagation through the full module hierarchy, focusing on pipeline depth timing, cross-module consistency, and X-propagation detection from broken submodules. Representative cases:

\begin{lstlisting}[basicstyle=\ttfamily\scriptsize,language=Verilog, frame=single]
// TC-INT-01: Cold start -> full pipeline settles, no X
wait(PIPELINE_DEPTH + MARGIN); check("TC-INT-01", top_output !== X);

// TC-INT-02: Spec-compliant inputs -> end-to-end no X
set(all inputs=<spec_values>); wait(PIPELINE_DEPTH ...
\end{lstlisting}

This layer provides the simulation-based pass/fail signal determining whether the assembled top-level design requires further iteration.

\paragraph{{\textbf{Step 4 — Testbench Assembly and Coverage Verification}}}
All submodule and integration testbenches are consolidated into a single unified file with independent top-level test modules per design, each producing parseable PASS/FAIL output consumed by an automated shell script for compilation, simulation, and aggregation. Coverage is verified by mapping every normative spec clause to at least one test case—a spec-clause standard stronger than code-path coverage.

\begin{table}[h]
\centering
\begin{tabularx}{\linewidth}{|X|X|} 
\hline
\textbf{Spec Clause} & \textbf{Derived Test} \\ \hline
"If input X is all-zero, output Y must equal the identity value." & \texttt{set(X=0); wait(N); check("TC-MOD-0N", Y === );} \\ \hline
\end{tabularx}
\end{table}

\paragraph{{\textbf{Step 5 - Synthesis and PPA Evaluation}}}
Beyond simulation, the Verification Agent also performs synthesis-based validation. Yosys\cite{Yosys} is used as the primary synthesis check, confirming that the generated RTL is synthesizable and mapping it to a gate-level netlist. OpenLane\cite{openlane2_woset2024} then runs the full physical design flow, performing broad design rule checks and reporting PPA—covering cell area, critical path delay, and power consumption—that quantify the quality of the generated RTL against standard physical implementation targets.

\begin{algorithm}[t]
\footnotesize
\captionsetup{font=footnotesize}
\caption{Progressive Coding Module}
\label{alg:coding_pipeline}
\KwIn{Knowledge Graph (HDA) $\mathcal{G} = \langle \mathcal{M}, \mathcal{E} \rangle$, Implementation Plan $\Pi$, max retries $R$}
\KwOut{Synthesizable Verilog $\{V_m\}$ for all $m \in \mathcal{M}$}

\BlankLine
\textcolor{red}{\tcp{Preprocessing}}
$\hat{\mathcal{M}} \leftarrow \textsc{SortDescTier}(\mathcal{M})$ \tcp*{leaf modules (highest tier) first}
$\Lambda \leftarrow \textsc{BuildIndex}(\Pi)$ \tcp*{map: $m.\mathrm{id} \mapsto \pi\text{-entry}$}

\BlankLine
\textcolor{red}{\tcp{Phase 1: Progressive Sub-module Generation}}
\For{each $m \in \{ m' \in \hat{\mathcal{M}} \mid \tau(m') > 0 \}$}{
    $P_m \leftarrow \textsc{PseudoCoder}\!\left(\nu(m),\; \Lambda[m]\right)$ \tcp*{$\nu(m)$: KG ports/params/constraints\; $\Lambda[m]$: plan functionality + steps}
    $\pi \leftarrow \textsc{PromptBuild}(P_m,\; \nu(m))$\;
    $V \leftarrow \textsc{Coder}(\pi)$\;
    $H \leftarrow [\,]$;\quad $t \leftarrow 1$\;
    \While{$t \leq R$}{
        $\langle \mathit{ok},\; E \rangle \leftarrow \textsc{SyntaxCheck}(V)$ \tcp{compilation}
        \lIf{$\mathit{ok}$}{\textbf{break}}
        $H \mathrel{+}= \langle t,\; \hat{E}_t \rangle$ \tcp*{$\hat{E}_t$: erroneous lines + error messages only}
        $\pi \leftarrow \textsc{PromptOptimizer}\!\left(\pi,\; V,\; E,\; H_{1:t-1}\right)$ \tcp*{latest $(V,\!E)$ in full\; $H_{1:t-1}$ compressed}
        $V \leftarrow \textsc{Coder}(\pi,\; V,\; E)$\;
        $t \leftarrow t + 1$\;
    }
    Save $\langle m, V \rangle$ as verified, or unverified if $t > R$\;
}

\BlankLine
\textcolor{red}{\tcp{Phase 2: Top-level Wrapper Assembly}}
\For{each $m \in \{ m' \in \hat{\mathcal{M}} \mid \tau(m') = 0 \}$}{
    $\mathcal{C} \leftarrow \bigl\{ m' \mid (m',\, m) \in \mathcal{E}_{\mathrm{inst}} \bigr\} \cap \mathcal{V}_{\mathrm{ok}}$ \tcp*{verified children via KG \texttt{instance\_of} edges}
    $W \leftarrow \textsc{CodeAssembler}\!\left(m,\; \mathcal{C},\; \mathcal{E}_{\mathrm{conn}}\right)$ \tcp*{parent spec + child Verilog + KG \texttt{connects} edges}
    Syntax-check $W$; retry up to $R_w$ times with error feedback\;
    Save $\langle m, W \rangle$\;
}
\end{algorithm}


%% file: experiments.tex
\section{Case Study: RISC-V 32I Processor}

We build a RISC-V 32I as a case study to demonstrate the implementation details of our framework.

\subsection{Case Study: Agent Pipeline for RV32I Spec Processing}

To demonstrate the operation of the proposed pipeline, we conducted a case study using the RV32I specification. The specification document was obtained from the official RISC-V user-level ISA documentation\cite{Waterman:EECS-2016-118}. Since hardware specifications are typically distributed as long PDF containing both textual descriptions and figures, the first step involves extracting structured information suitable for LLM processing.

The specification text was extracted using PyPDF, while figures and tables referenced in the document were automatically captured using a custom preprocessing tool. These visual artifacts are important because architectural specifications often encode critical information through diagrams, dataflow representations, and lookup tables. The extracted content is then processed by a set of coordinated LLM agents within the Architectural Analysis Module. We have used Autogen\cite{wu2024autogen} framework for our multi-agent pipeline.

Following preprocessing, the extracted specification content is processed by the multi-agent pipeline implemented in the Architectural Analysis Module. All agents in this stage were implemented using the \textit{GPT-4o-mini}\cite{openai2024gpt4omini} model to balance reasoning capability and computational efficiency during iterative analysis of large specification documents.

The first stage of the pipeline performs section-level summarization of the specification. Each section of the document is analyzed and converted into a structured JSON summary containing the section purpose, key technical concepts, important terminology, and cross-section dependencies. In total, the pipeline processed \textbf{88 sections} inside \textbf{238 pages} of the RISC-V specification, producing structured summaries that provide an LLM-friendly representation of the architecture while preserving implementation-relevant details. These summaries serve as the contextual foundation for downstream architectural decomposition. We filter out irrelevant sections from the summaries for the decomposition to save input tokens.

Using both the summaries and the original specification content, the decomposition and description stages generate a structured \textbf{implementation plan} representing the RV32I architecture as a collection of modular hardware components. During this stage, the content auditor agent evaluates the generated module descriptions to ensure completeness, correctness, and readiness for RTL implementation. The final output of this process is a structured JSON representation describing each module, its interface, functionality, and implementation constraints. An excerpt of the generated implementation plan is shown below:

\begin{lstlisting}[basicstyle=\ttfamily\scriptsize,caption={Simplified excerpt of the Implementation Plan generated by Architectural Analysis Module}]
{
  "target_function": "RISC V 32I",
  "implementation_plan": [
    {
      "name": "RegisterFile",
      "description": "Implements a register file with 32 general-purpose registers (x0 to x31) and a program counter (pc).",
      "inputs": [
        "read_addr1 (5-bit)",
        "read_addr2 (5-bit)"...
      ],
      "outputs": [
        "read_data1 (32-bit)",...
      ],
      "functionality": "The RegisterFile stores 32 general-purpose registers and a program counter, allowing for read and write operations based on specified addresses...",
      "steps": [
        "1. Initialize a 32-entry array of 32-bit registers to store the register values.",
        "2. Set register x0 to zero during initialization, ensuring it always returns zero on read....
      ],
      "references": [
        "Section 2.1",
        "Figure 2.1"
      ],
      "implementation_notes": "Ensure that the register array is indexed correctly, with x0 hardwired to zero. The ...",
      "order": 1,
      "dependencies": []
    },
    {
      "name": "InstructionDecoder",
      "description": "Decodes incoming instructions into their respective formats (R-type, I-type, S-type, U-type, B-type, J-type).",
      "inputs": [
        "instruction (32-bit)"...
\end{lstlisting}

\begin{figure}[htbp] 
    \centering
    \includegraphics[width=1\linewidth]{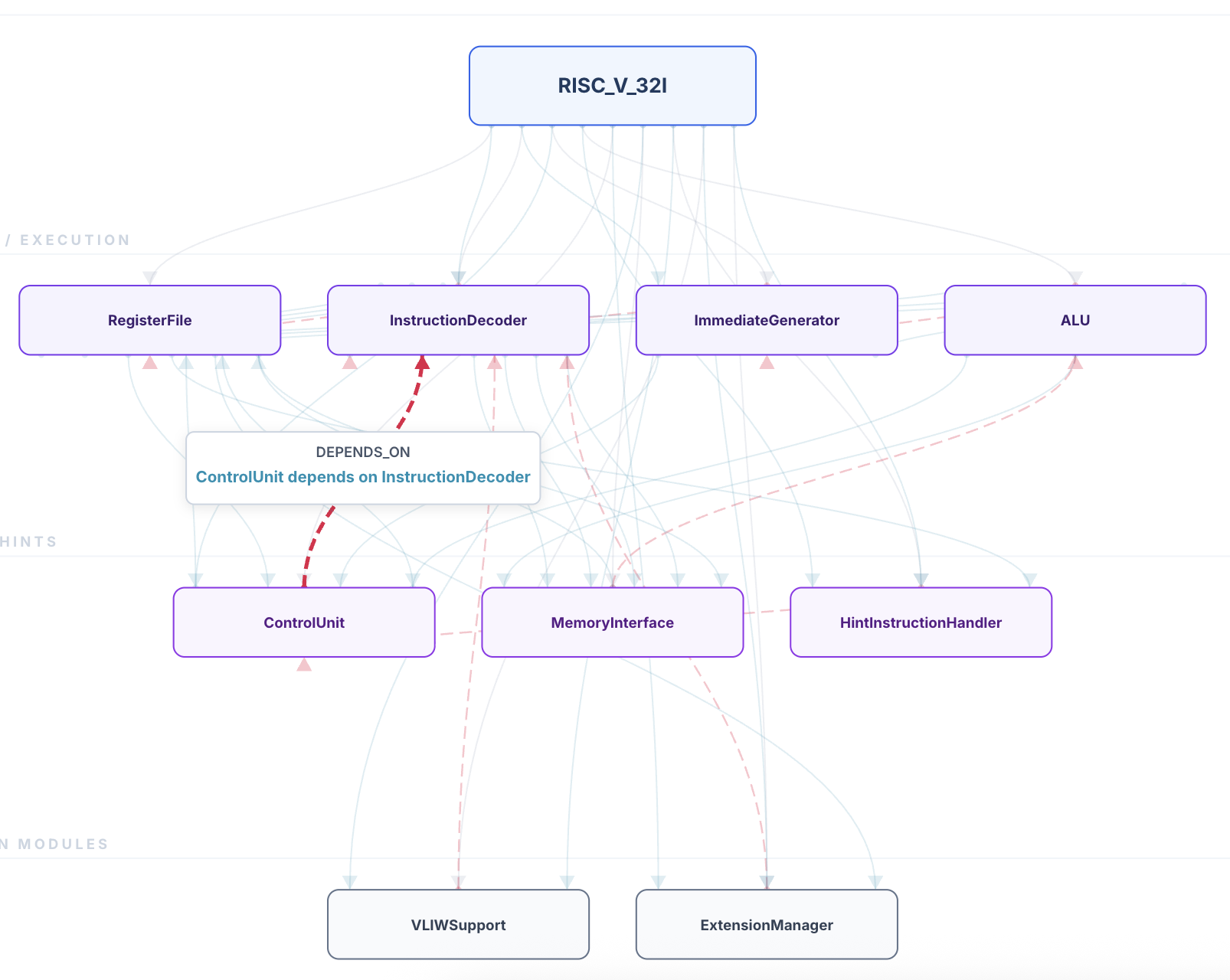} 
    \caption{Hierarchical Design
Architecture (HDA) for RV32I.}
    \label{fig:hda_ovr}
\end{figure}

This structured implementation plan is then used by the \textit{KG Builder Agent} to construct the \textbf{Hierarchical Design Architecture (HDA)}, a machine-readable knowledge graph whose nodes represent hardware modules and whose edges capture hierarchical and wiring relationships. For RV32I, these nodes include modules such as \texttt{RISC\_V\_32I}, \texttt{RegisterFile}, \texttt{InstructionDecoder}, \texttt{ImmediateGenerator}, \texttt{ALU}, \texttt{ControlUnit}, \texttt{MemoryInterface}, \texttt{HintInstructionHandler}, \texttt{ExtensionManager}, and \texttt{VLIWSupport}. Fig.~\ref{fig:hda_ovr} shows the resulting RV32I graph, illustrating how major components are connected through explicit structural relations. The generated HDA is stored in JSON format and visualized through an interactive static website.\footnote{The complete interactive knowledge graph and detailed data structures can be accessed at: \url{https://sazzadsowmik.github.io/risc-v-32I-KG-demo/}}

This graph representation serves as the structural backbone of the pipeline.

Given the structured implementation plan and the Hierarchical Design Architecture (HDA), the Progressive Coding Module generates RTL in a staged manner. For the RV32I case study, the Pseudo Coder Agent first produces pseudo-code for each submodule. An example pseudo-code for the ALU is shown below:

\begin{lstlisting}[basicstyle=\ttfamily\scriptsize,language=Verilog, caption={Pseudo-code for ALU generated by Pseudo Coder Agent}]
MODULE ArithmeticLogicUnit (
  INPUT: operand1 [31:0], operand2 [31:0], alu_control [3:0],
  OUTPUT: result [31:0], zero_flag [0:0], overflow_flag [0:0]
)

  REGISTERS:
    reg [31:0] result           // Result of the ALU operation
    reg [0:0] zero_flag         // Zero flag output
    reg [0:0] overflow_flag      // Overflow flag output

  INITIALIZE:
    result = 32'b0              // Initialize result to zero
    ...

ALWAYS @(operand1, operand2, alu_control):
  // Step 1: Decode alu_control to determine operation
  CASE (alu_control):
    4'b0000:  // ADD operation
      result = operand1 + operand2
      overflow_flag = (operand1[31] == operand2[31]) && (result[31] != operand1[31])  // Check for overflow
    4'b0001:  // SUB operation
    .
    .
    .
    DEFAULT:
      result = 32'b0  // Default case, result is zero
      overflow_flag = 1'b0  // No overflow for default case
  // Step 6: Set zero_flag if result is zero
  zero_flag = (result == 32'b0) ? 1'b1 : 1'b0
END MODULE
\end{lstlisting}

\textit{The Coder Agent} then translates the pseudo-code into synthesizable RTL. It is guided to generate clean, synthesizer-friendly Verilog code that adheres to standard coding practices and avoids tool-specific incompatibilities. The generated RTL is intended to be compatible with downstream compilation using Icarus Verilog and synthesis flows such as Yosys\cite{Yosys} and OpenLane\cite{openlane2_woset2024} for PPA evaluation.

\textbf{Iterative Refinement.}
Each generated submodule is validated using the \textit{Syntax Checker Agent}. If compilation fails, the \textit{Prompt Enhancer Agent} refines the generation using error feedback. As described in Section \ref{subsec:framework_progressive_coding_module}), only the latest full code and error logs are retained, while earlier iterations are summarized to reduce token usage.

\textbf{Cost Analysis.}
We further analyze the computational cost of the RV32I case study.

\begin{lstlisting}[basicstyle=\ttfamily\scriptsize,caption={Cost summary (RV32I pipeline)}]
Architectural Analysis    : $0.7174
Hierarchy Analysis Module : Negligible (skipped)
Progressive Coding        : $0.0265
Total                     : $0.7439
\end{lstlisting}

\textbf{Functional Correctness.}
We evaluate functional correctness using specification-driven testbenches applied to the RV32I design. The generated processor successfully passes all 183 test cases, including both submodule-level validation and randomized arithmetic tests for the ALU, demonstrating correct behavioral execution across key components.

\subsection{Case Study: Other experiments}
We also leverage three widely recognized cryptographic standards from the NIST Federal Information Processing Standards (FIPS): the Advanced Encryption Standard (AES)\cite{nist_aes}, the Digital Signature Standard (DSS)\cite{nist_dss}, and the Keyed-Hash Message Authentication Code (HMAC)\cite{nist_hmac}. By utilizing these diverse benchmarks, we provide an exhaustive evaluation of our tool across a variety of complex hardware design tasks.
Their specification documents vary in size and structure, spanning 52 pages with 5 sections, 15 figures, and 8 tables for AES, 86 pages with 7 sections, 4 figures, and 4 tables for DSS, and 13 pages with 6 sections, 1 figure, and 1 table for HMAC.

These results demonstrate that the staged generation approach not only produces structurally consistent RTL but also maintains low computational cost through efficient iterative refinement. The code will be open on GitHub upon publication.

%% file: results.tex
\section{Experimental Results}

\underline{Evaluation Metrics.}
We evaluate pipeline effectiveness using two metrics: (1) \textbf{Intervention}, defined as the total number of human fixes required across both simulation and synthesis to achieve syntactically correct RTL — counted per module and reported as the highest fix count observed across all modules in the design; and (2) \textbf{Coding}, defined as the average number of code generation and revision iterations required per submodule, capturing the efficiency and convergence behavior of the \textit{Progressive Coding Module}. Note that Spec2RTL-Agent\cite{yu2025spec2rtl} does not report intervention at per-module granularity; for a fair comparison we align our reported intervention figure to their definition by taking the per-module maximum.

\begin{table}[h]
    \centering
    \scriptsize
    \captionsetup{font=scriptsize}
    \caption{\textsc{Performance Comparison: VeriGraphi vs. Spec2RTL\cite{yu2025spec2rtl} (Lower is better).}}
    \label{tab:performance_comparison}
    \begin{tabular}{lcccccccc}
        \toprule
        & \multicolumn{2}{c}{RISC-V-32I} & \multicolumn{2}{c}{DSS} & \multicolumn{2}{c}{HMAC} & \multicolumn{2}{c}{AES} \\
        \cmidrule(lr){2-3} \cmidrule(lr){4-5} \cmidrule(lr){6-7} \cmidrule(lr){8-9}
        Metric & VG & S2R & VG & S2R & VG & S2R & VG & S2R \\ 
        \midrule
        Intervention & \textbf{2} & -- & 6 & 6 & \textbf{0} & 3 & 8 & \textbf{4} \\
        Coding       & \textbf{1.11} & -- & \textbf{2.60} & 9.31 & \textbf{1.14} & 9.52 & \textbf{3.00} & 8.49 \\
        \bottomrule
        \addlinespace
        \multicolumn{9}{l}{\scriptsize VG: VeriGraphi (Ours), S2R: Spec2RTL-Agent}
    \end{tabular}
\end{table}

\underline{Synthesizability Validation.}
 We first use Yosys to confirm that the generated RTL is synthesizable. For designs that pass this check, we run the full OpenLane flow, which performs RTL-to-layout implementation using Yosys/OpenROAD-based tools under the SkyWater open PDK \texttt{sky130A} (nominal 130 nm). The results in Table\ref{tab:synthesis_results} formally confirm that VeriGraphi-generated RTL is fully synthesizable through this standard physical implementation flow, with reported area, cell count, critical path, and power metrics obtained from the final flow.

 \begin{table}[h]
    \centering
    \scriptsize
    \setlength{\tabcolsep}{6pt} 
    \captionsetup{font=scriptsize}
    \caption{\textsc{Synthesis Results and Physical Metrics.}}
    \label{tab:synthesis_results}
    \begin{tabular}{lccccc}
        \toprule
        Design & \begin{tabular}[c]{@{}c@{}}Area\\ (mm²)\end{tabular} & \begin{tabular}[c]{@{}c@{}}Total\\ cells\end{tabular} & \begin{tabular}[c]{@{}c@{}}Clk\\ (ns)\end{tabular} & \begin{tabular}[c]{@{}c@{}}CP\\ (ns)\end{tabular} & \begin{tabular}[c]{@{}c@{}}Power\\ (mW)\end{tabular} \\
        \midrule
        RISC-V32I  & 0.3600 & 35,867 & 10.0 & 4.40 & 0.9000268 \\
        HMAC       & 0.0784 & 7,168  & 10.0 & 1.39 & 1.4110073 \\
        DSS        & 0.0784 & 6,933  & 10.0 & 1.23  & 4.9500000      \\
        AES        & 4.0000 & 481,941  & 40.0 & 32.1 & 5,687.1020 \\
        \bottomrule
        \addlinespace
        \multicolumn{6}{l}{\scriptsize CP: Critical Path.}
    \end{tabular}
\end{table}

%% file: conclusion_futureWork.tex
\section{Conclusion and Future Works}

In this work, we presented a structured, knowledge graph-driven pipeline for translating unstructured specification documents into synthesizable RTL. By introducing a spec-anchored Hierarchical Design Architecture and a progressive multi-agent generation flow, VeriGraphi improves structural consistency, reduces hallucinated interfaces, and enables reliable generation of hierarchical Verilog designs. We further validate the generated RTL through specification-driven functional testing, synthesis, and PPA evaluation, and show successful generation of complex designs such as RISC-V with minimal human intervention and HMAC with zero intervention, while requiring substantially fewer coding iterations than prior work. The hierarchical structure also brings practical efficiency benefits: local errors can be corrected without regenerating unrelated modules, and intermediate outputs from each stage are preserved as checkpoints, avoiding restarts from the beginning of the pipeline.

As future work, we plan to extend VeriGraphi in three directions. First, we aim to introduce a \textit{Global Code Library} as a unified graph-based memory for reusable verified modules across designs. Second, we will evaluate the framework on more complex hierarchical designs beyond RISC-V. Third, we plan to add extra constraints and optimization objectives to improve generation quality and implementation results.

%% file: acknowledgement.tex
\section{Acknowledgement}

We acknowledge the assistance of Marwan Abdelwahab in streamlining the synthesis and evaluation workflow. 

This work is supported by the National Science Foundation under Grant No. 2434247. Any opinions, findings, and conclusions or recommendations expressed in this material are those of the author(s) and do not necessarily reflect the views of the funding agencies.